\documentclass[5p,preprint]{elsarticle}

\usepackage{hyperref}
\usepackage{gensymb}
\usepackage{chemfig}
\usepackage[version=3]{mhchem}
\setatomsep{1em}
\journal{Cryogenics}

\begin{document}

\begin{frontmatter}

\title{Specific heat, thermal conductivity, and magnetic susceptibility of cyanate ester resins --- An alternative to commonly used epoxy resins}

\author[crc]{Sachiko Nakamura\corref{mycorrespondingauthor}}
\cortext[mycorrespondingauthor]{Corresponding author}
\ead{snakamura@crc.u-tokyo.ac.jp}
\author[crc]{Takenori Fujii}
\author[mgc]{Shoji Matsukawa\fnref{mgca}}
\author[mgc]{Masayuki Katagiri}
\author[crc,sphys]{Hiroshi Fukuyama}

\address[crc]{Cryogenic Research Center, The University of Tokyo, 2-11-16, Yayoi, Bunkyo-ku, Tokyo 133-0032, Japan}
\address[mgc]{Mitsubishi Gas Chemical Company, Inc., 2-5-2, Marunouchi, Chiyoda-ku, Tokyo 100-8324, Japan}
\fntext[mgca]{Present address: Mitsubishi Gas Chemical America, Inc., 655 Third Avenue, 19th Floor, New York, NY 10017.}
\address[sphys]{Department of Physics, The University of Tokyo, 7-3-1, Hongo, Bunkyo-ku, Tokyo 133-0033, Japan}

\begin{abstract}
In low temperature experiments, resins have many applications as glues or thermal and electrical insulators. 
Cyanate ester resins (CEs) are a high-temperature compatible thermoset resin whose glass-transition temperature $T_\mathrm{g}$ is $\approx300$\,\degree C. 
Recently, we found that CEs also withstand low temperatures without microcracking by measuring $^4$He permeability. 
Here, we measured specific heat $C$, thermal conductivity $\kappa$, and magnetic susceptibility $\chi$ of different kinds of CEs in the wide temperature range from room temperature to 0.5\,K for $C$ and 2\,K for other two. 
The thermal properties, $C$ and $\kappa$, of different kinds of CEs are surprisingly coincident with each other. 
We discuss chemical structures and crystallinity of CEs and their blends based on the measured thermal properties. 
Compared to Stycast 1266, a commonly-used epoxy resin in low temperature experiments,  $C$ of CEs is larger by a factor of 3 ($\leq30$\,K), $\kappa$ is lower by a factor of 4 ($\leq10$\,K), indicating the small thermal diffusivity. 
The $\chi$ values are as small as Stycast 1266, indicative of their high purity. 
Our results show that cyanate esters are a new option for cryogenic resins with thermal insulative properties in/for low temperature experiments. 
\end{abstract}

\end{frontmatter}


\section{Introduction}
Cyanate ester resins (CEs) are a strong and stiff thermoset resin, which offers advantages as a matrix resin of FRPs (fiber reinforced composites) because of their high thermal stability, low water absorption, low outgassing properties, small dielectric loss, and high radiation resistance~\cite{CTCER}. 
Since the late 1970s, FRPs of CEs or their copolymers have widely been used in the electronic and aerospace industries under severe environmental conditions such as extremely high/low temperature and/or in vacuum. 
They are also attracting attention for their excellent cryogenic durability to microcracking~\cite{doi:10.1063/1.2192356} and additional functions as a thermal management material~\cite{SHINOZAKI2015270} in space applications.  

On the other hand, neat resins of CEs (without fibers or fillers) are not commonly used at cryogenic temperature. 
As a casting resin or glue for low temperature, epoxy resins (EPs) are widely used so that we can access the vast database of their cryogenic properties~\cite{pobell,0022-3719-14-12-007,scheibner1985thermal}. 
Unlike EPs, limited measurements have been reported for CEs at low temperature~\cite{Walsh1994}. 
To our knowledge, even the thermal conductivity or specific heat have not been measured yet. 

In principle, CEs can take the place of EPs in most applications because they share a number of pre- and post-cure qualities with common EPs. 
CEs have the following features which have benefits over EPs and thus could see new applications: higher glass-transition temperature ($T_\mathrm{g}\approx 300$\,\degree C), lower dielectric loss, and absence of irritant or allergic reactions~\cite{CTCER}. 
Recently, we suggest that CEs are an excellent alternative to EPs especially in surface-sensitive experiments at low temperature based on our measurements of $^4$He permeation properties at 77--340\,K and H$_2$O uptake at room temperature ~\cite{LT28_CE}. 
However, the lack of vast materials database of various kinds of CEs on much more common thermal and magnetic properties, such as $\kappa$, $C$, and magnetic susceptibility $\chi$, particularly below room temperature still discourages the broad use of CEs in low temperature experiments. 
Especially, $C$ is important to verify the existence/absence of structural transitions which may cause drastic change in other material properties, along with its technical importance to estimate heating/cooling time required to change the temperature. 

Here, we measured the $C$ (0.5--300\,K), $\kappa$ (2--300\,K), and $\chi$ (2--400\,K) of several kinds of CEs at wide temperature ranges indicated in the parentheses. 
We also discuss the crystallinity and chemical structures based on amorphous contributions in the $C$ and $\kappa$ below 10\,K, and impurity concentrations based on diamagnetic and paramagnetic components of the $\chi$.

\section{Experiments}
\subsection{Sample details}
\begin{figure}[tb]
\centering
\includegraphics[width=0.9\columnwidth]{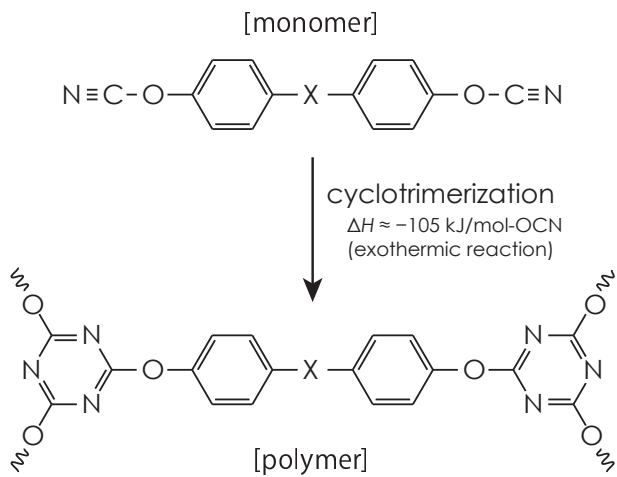}
\caption{Typical chemical structures of monomer and polymer of aromatic cyanate ester resins.}
\label{chem}
\end{figure}

Cyanate monomers contain multiple \ce{-O-C#N} groups. 
After repeating cyclotrimerization of three \ce{-O-C#N} groups, rigid triazine rings form a strong three-dimensional polymer network. 
Typical chemical structures of the monomer and polymer are illustrated in Fig.~\ref{chem}.

\paragraph{Cyanate esters}
Three kinds of CE monomers~\cite{mgc} were investigated: 
TA (2,2-bis(4-cyanatophenyl)propane), 
P-201 (1,1-bis(4-cyanatophenyl)ethane), 
and NCN (aralkyl type cyanate ester resin). 
At room temperature, the appearance of each CE monomer varies dramatically. 
The monomer of TA is white flakes, P-201 is a yellow liquid, and NCN is a brown solid which is treated as a clear organic solution. 
The monomers are pretreated by melting them at 120\,\degree C with stirring and having them degass in a vacuum.
The CE can then be poured into a mold made of PTFE or surface-treated metals. 
As the melt viscosity of TA and P-201 is very low ($\approx 10$\,mPa$\cdot$s at 100\,\degree C), they are suitable for copying fine structures of the mold. 
Typical curing procedure is 150\,\degree C for 3 hours, 180\,\degree C for 5 hours, then 250\,\degree C for 5 hours in ambient pressure. 
The stepwise heating is important to prevent runaway reactions due to the topical exothermic polymerization. 
The cured resins (polymers) are a hard solid whose colors are clear yellow for TA, clear amber for P-201, black for NCN, and intermediate colors for their mixtures. 
The densities measured at room temperature are 1.17(5)\,g/cc for TA, 1.20(7)\,g/cc for P-201, 1.19(2)\,g/cc for NCN, and 1.17(1)\,g/cc for copolymer of TA and NCN. 

As there are no byproducts from the polymerization, CE polymers rarely include voids, except NCN including the vaporized solvent. 
To avoid void formation in the NCN sample for thermal conductivity measurements, the first step of the heating at 150\,\degree C was performed in vacuum (at 80\,kPa) to pump out the solvent fume. 
The surface color and void structure varied with curing conditions, but no significant effects on the properties of the solid portion were found when measuring $C$, $\kappa$, and $\chi$. 

\paragraph{BT resins}
BT resins are blends of CEs and bismaleimides (BMIs)~\cite{SatoshiAYANO1984}. 
We prepared two types of BT resins: 
BT1 is a mixture of TA and BMI70 (bis-(3-ethyl-5-methyl-4-maleimidephenyl)methane),  
while BT2 is of TA and BMI2300 (phenylmethane maleimide~\cite{daiwakasei}). 
Both were cured in similar conditions to those for CEs. 
The mixing ratio of TA and BMI is 3:2 by weight. 
The densities measured at room temperature are 1.14(2)\,g/cc for BT1 and 1.15(5)\,g/cc for BT2.

\subsection{Magnetization measurements}
$\chi$ of CEs (2 samples of TA and one for each of P-201, NCN, and copolymer of TA and NCN), BT resins (BT1, BT2), and Stycast 1266 were measured using Magnetic Property Measurement System (MPMS)~\cite{qd} with Reciprocating Sample Option (RSO). 
The samples were 30--50\,mg in weight. 
In the measurements, the samples were inserted to a pair of slits in a straw. 
The straw has a positive and small $\chi$ (5--7\% of the total signal), which depends very little on the temperature.  
The $\chi$ of the straw is subtracted in the results. 

\subsection{Specific heat measurements}
The specific heats of the CE samples were measured using Heat Capacity Option of  Physical Property Measurement System (PPMS)~\cite{qd} at temperature ranges 1.8--300\,K and 500\,mK--50\,K (with helium-3 refrigerator). 
The samples are square plates of $\approx4$\,mm side and $\approx1$\,mm in thickness, 10--20\,mg in weight. 
The samples were loaded on the alumina platform of the sample pack using GE 7031 varnish~\cite{ge}. 

\subsection{Thermal conductivity measurements}
Thermal conductivity of the CE samples was measured using Thermal Transport Option of PPMS~\cite{qd} at temperatures between 2\,K and 300\,K. 
We mainly used four-terminal method to minimize effects of the contact resistance between the sample and the gold-plated copper leads independently glued to the sample with Stycast 2850FT. 
The samples were shaped as 4-mm square rod of 12\,mm in length, whose thermal relaxation time is too long at higher temperature ($T>$100--200\,K). 
Thus a thin sample of P-201 (5-mm diameter and 2\,mm-thick cylinder) was also measured using two-terminal method, and it was found that the data of two samples overlap within 5\% discrepancy at temperatures from 4 to 100\,K. 
The effects of the contact resistance in two-terminal method appears at lower temperature. 

\section{Results and discussions}
\subsection{Magnetic susceptibility}
Polymers are often used in magnetic fields because normal insulative polymers do not usually have strong ferromagnetic properties. 
However, when we use them in very strong magnetic fields ($H\geq10$\,kOe~\cite{polymerjournalKimura2003}), we should screen the polymer products because coloring agent, catalysts, impurities, and curing residues can have strong magnetic properties. 
Stycast 1266~\cite{stycast} is known to have a small magnetic susceptibility $\chi$~\cite{doi:10.1063/1.1137340}, which is one order smaller than other polymers~\cite{WAPLER2014233}.

\begin{figure}[b!]
\centering
\includegraphics[width=0.9\columnwidth]{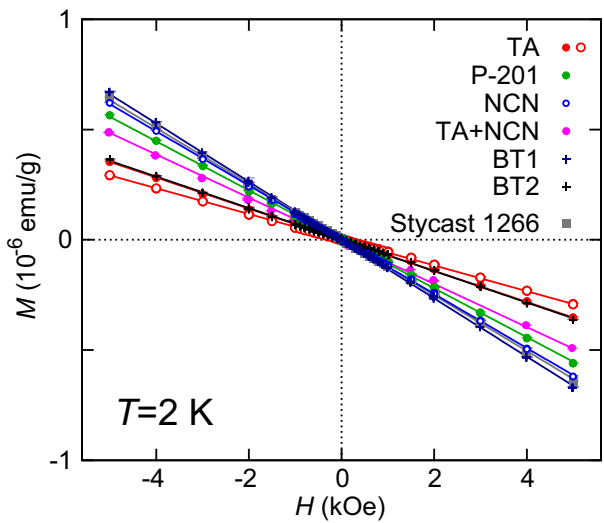}
\caption{Magnetization at 2\,K of CE (TA, P-201, NCN, TA+NCN), BT resins (BT1, BT2), and EP (Stycast 1266). Error bars are much smaller than the symbols.}
\label{CE_MH}
\end{figure}

$M$-$H$ curves (magnetization $M$ versus magnetic field $H$) were measured from $H=-5$ to 5\,kOe at $T=2$\,K and 300\,K. 
They are a featureless straight line for each sample at each temperature as shown in Fig.~\ref{CE_MH}. 
Thus, $\chi$ was calculated from the $M$ measured with a fixed $H$ ($=5$\,kOe) at temperatures between 2 and 400\,K. 

\begin{figure}[t]
\centering
\includegraphics[width=0.92\columnwidth]{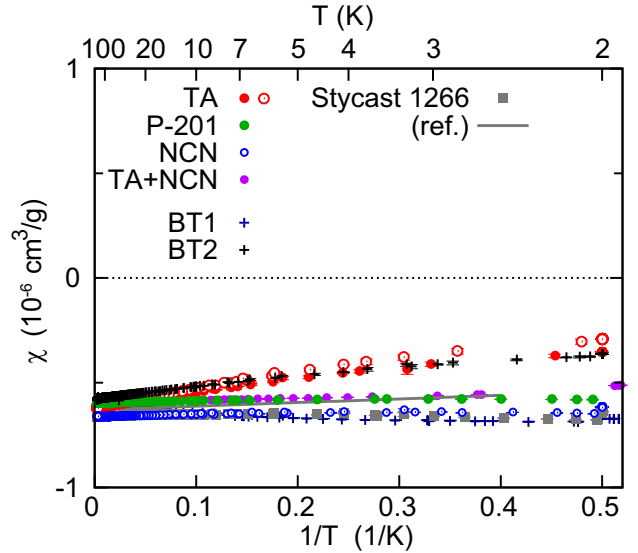}
\caption{Magnetic susceptibility at 5\,kOe of CE (TA, P-201, NCN, TA+NCN), BT resins (BT1, BT2), and EP (Stycast 1266; the line shows data from Ref.~\citenum{doi:10.1063/1.1137340}). Error bars are smaller than the symbols.}
\label{CE_Chi_at_5kOe}
\end{figure}

The temperature dependence of $\chi$ is shown in Fig.~\ref{CE_Chi_at_5kOe}. 
The $\chi$ values are very small ($|\chi|<1\times10^{-6}$\,cm$^{3}$/g) and this polymer family is a potential material to use in high magnetic fields over wide temperature ranges. 
At room temperature, CEs and BT resins have similar negative values of $\chi$ to Stycast 1266. 
Along with this diamagnetic contribution of $\approx -6\times10^{-7}$\,cm$^3$/g (the intercept in Fig.~\ref{CE_Chi_at_5kOe}), the paramagnetic component (the slope in Fig.~\ref{CE_Chi_at_5kOe}) becomes distinguishable at low temperature especially for TA and BT2. 
The Curie constants correspond to free spin densities ($g=2$) of $1.5\times10^{17}$--$4.4\times10^{18}$ per gram. 
As the CEs include $\approx 2\times10^{21}$ triazine rings per gram, it indicates that free spins are found only once or twice for a thousand of triazine rings. 
The paramagnetic components of P-201, NCN, copolymer of TA and NCN, and BT1 are as small as Stycast 1266. 

\subsection{Specific heat}\label{sec_sh}
\begin{figure}[bt]
\centering
\includegraphics[width=0.97\columnwidth]{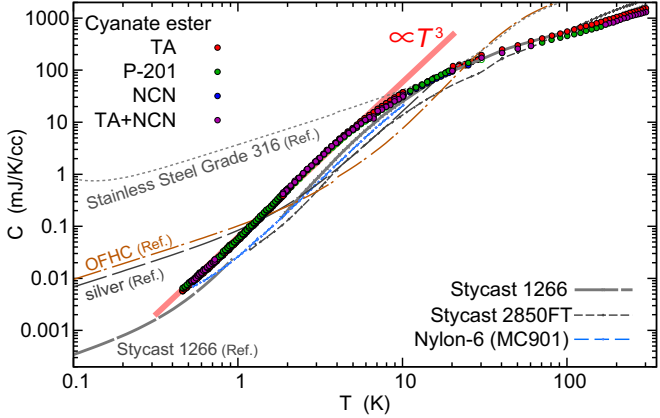}
\caption{Volumetric specific heats of CEs (TA, P-201, NCN, and copolymer of TA and NCN), EP (Stycast 1266), EP with filler (Stycast 2850FT), and Nylon-6 (MC901). Measurement errors are typically 1\%, which is much smaller than the symbols. Stainless Steel Grade 316~\cite{nist}, OFHC~\cite{nist}, silver~\cite{pobell}, and lower temperature data of Stycast 1266~\cite{pobell} in previous studies are also shown for comparison.}
\label{TA_HC_JPS2_2}
\end{figure}

The volumetric specific heats $C$ measured for CEs (TA, P-201, NCN, and copolymer of TA and NCN~\cite{mgc}), EP (Stycast 1266~\cite{stycast}), EP with filler (Stycast 2850FT~\cite{stycast}), and Nylon-6 (MC901~\cite{mc901}) are plotted in Fig.~\ref{TA_HC_JPS2_2}, along with those of metals and Stycast 1266 derived from previous studies~\cite{pobell,nist}. 
Here, the thermal contraction is ignored as it is less than 1\%~\cite{Walsh1994}.  
Surprisingly, all CE samples have almost the same $C$. 
The difference between them is less than 5\% at $T<30$\,K and $\approx$26\% at most. 
At temperatures between 30 and 300\,K, CEs have similar $C$ to other synthetic resins, which is much smaller than that of metals. 
At low temperatures ($T<20$\,K), CEs have larger $C$ than other synthetic resins do.
$C$ of CE is smaller than that of Stainless Steel Grade 316 alloy below $\approx 10$\,K. 
At $T<5$\,K, the $C$ shows $\approx T^3$ behavior expected for phonons in three dimensions.  
At $\approx1$\,K, CEs have much smaller $C$ than metals. 
Throughout the temperature range we studied, no specific heat anomalies indicative of phase transitions are found thus we can use CEs in low temperature devices without expecting drastic change in material properties. 

At $T<1$\,K, other polymers show an additional $T^n$ ($n\approx1$) contribution which appears to be a concave up in Fig.~\ref{TA_HC_JPS2_2}.  
This type of contribution is caused by tunneling transitions or structural relaxation which undergo when each atom has several possible positions distinguished by rather small energy~\cite{pobell}. 
The upturn is a characteristic behavior of noncrystalline or disordered solids not arranged in a periodic order, but it is hardly distinguishable in CEs even below 1\,K.  
This indicates that CEs have low possibility of performing structural rearrangements in atomic scales. 

\subsection{Thermal conductivity}
Resins are often used for heat insulation in low temperature experiments because of their low thermal conductivity $\kappa$. 
Along with the low thermal flow rate in equilibrium conditions, they are also used to isolate the system from rapid temperature fluctuation through the use of their low thermal diffusivity (the ratio between $\kappa$ and volumetric specific heat $C$). 

\begin{figure}[t!]
\centering
\includegraphics[width=0.95\columnwidth]{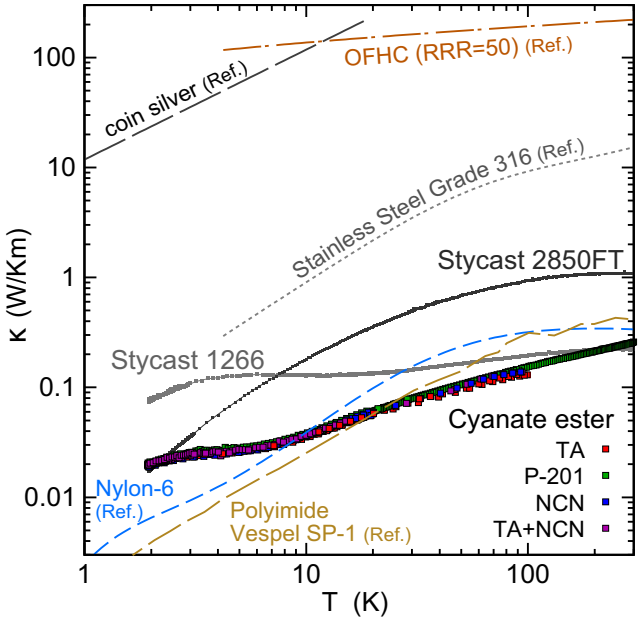}
\caption{Thermal conductivities of CEs (TA, P-201, NCN, and copolymer of TA and NCN), EP (Stycast 1266), and EP with filler (Stycast 2850FT). Measurement errors are typically 1\%, which is much smaller than the symbols. Stainless Steel Grade 316~\cite{nist}, OFHC~\cite{nist}, coin silver~\cite{pobell}, Nylon-6~\cite{pobell}, and polyimide (Vespel SP-1)~\cite{doi:10.1063/1.3292433} in previous studies are also shown for comparison.}
\label{CE_TT_jps2_2}
\end{figure}

The thermal conductivity $\kappa$ measured for CEs (TA, P-201, NCN, and copolymer of TA and NCN), EP (Stycast 1266), EP with filler (Stycast 2850FT) are plotted in Fig.~\ref{CE_TT_jps2_2}, along with those of polyimide (Vespel SP-1~\cite{kaptonvespel})~\cite{doi:10.1063/1.3292433}, Nylon-6, Stycast 1266, and metals reported in the previous studies~\cite{pobell,nist}. 
The difference in $\kappa$ between different kinds of CEs is again very small. 
The $\kappa$ is smaller than that of other materials at $T>20$\,K.  
At 2\,K, the lowest temperature in this study, the $\kappa$ of CEs are between low-$\kappa$ resins (such as Nylon and Vespel SP-1) and high-$\kappa$ resins (like Stycast 1266). 
Note that Stycast 2850FT has much smaller $\kappa$ than Stycast 1266 at $T\approx2$\,K.  
At $3<T<6$\,K, the $\kappa$ has a plateau in the $T$-dependence. 
Such plateaus are commonly found in amorphous materials (See Section~\ref{sec_pet}). 

\subsection{Thermal diffusivity}
\begin{figure}[t!]
\centering
\includegraphics[width=0.95\columnwidth]{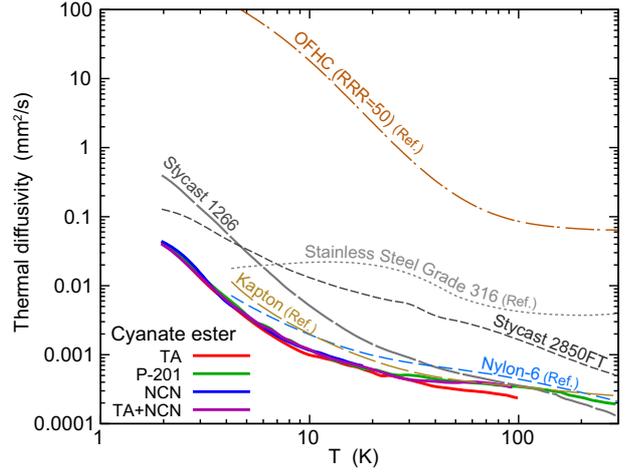}
\caption{Thermal diffusivity of CEs (TA, P-201, NCN, and copolymer of TA and NCN), EP (Stycast 1266), and EP with filler (Stycast 2850FT). Error bars are typically 1--2\%.  Stainless Steel Grade 316~\cite{nist}, OFHC~\cite{nist}, Nylon-6~\cite{nist}, and polyimide (Kapton~\cite{kaptonvespel})~\cite{nist} in previous studies are also shown for comparison.}
\label{TD}
\end{figure}

Comparing Figs.~\ref{TA_HC_JPS2_2} and \ref{CE_TT_jps2_2}, we find that large $C$ does not always lead to large $\kappa$. 
Figure~\ref{TD} shows the ratio of them ($\equiv \kappa/C$, known as thermal diffusivity) calculated from the $\kappa$ and $C$.  
Clearly, polymers have small values of $\kappa/C$ with CEs having the lowest. 
This indicates that CEs have an advantage to quench unwanted temperature fluctuations. 
As different kinds of CEs have almost the same thermal properties, the characteristic thermal properties are presumably due to the basic skeleton of CEs (folded triazine rings)~\cite{doi:10.1246/cl.2011.309}.  

\subsection{Relationship between crystallinity and thermal properties}\label{sec_pet}
To estimate the crystallinity of the CEs semi-quantitatively, we compare their thermal properties with those of polyethylene terephthalate (PET). 
PET is a thermoplastic polymer resin, which is a model material to study the contribution of amorphous structure because its crystallinity can be controlled by thermal history~\cite{ASSFALG19751389}. 

\begin{figure}[bt]
\centering
\includegraphics[width=0.97\columnwidth]{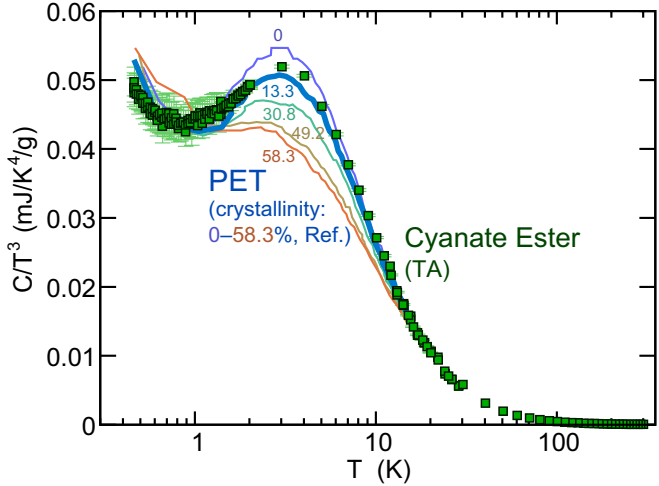}
\caption{Specific heat over temperature cube of CE (TA) compared to PET with varying degrees of crystallinity~\cite{0022-3719-20-20-006}. Error bars are smaller than the symbols at temperatures above 2\,K.}
\label{CE_HCcube_PET}
\end{figure}

\begin{figure}[bt]
\centering
\includegraphics[width=0.97\columnwidth]{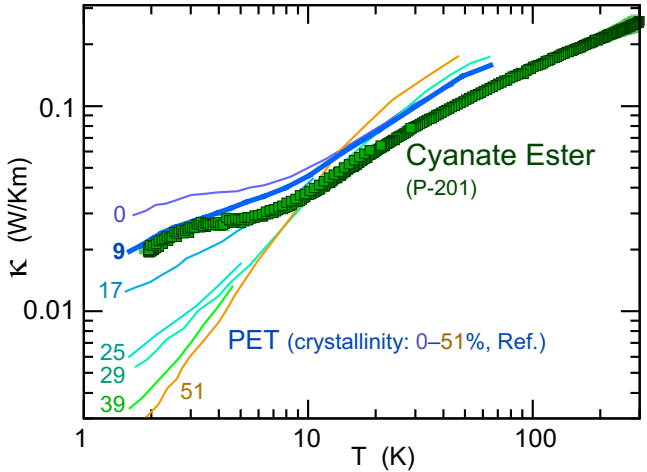}
\caption{Thermal conductivity of CE (P-201) compared to PET with varying degrees of crystallinity~\cite{GREIG1988243}. Error bars are smaller than the symbols at temperatures below 250\,K.}
\label{CE_TT_PET}
\end{figure}

$T$-dependences of $C$ and $\kappa$ are compared for CEs and PET with various degrees of crystallinity determined from small angle X-ray scattering~\cite{ASSFALG19751389} in Figs.~\ref{CE_HCcube_PET} and \ref{CE_TT_PET}, respectively~\cite{0022-3719-20-20-006,GREIG1988243}. 
At $T>15$\,K, crystalline PET has larger $\kappa$ than amorphous PET. 
This is because thermal phonon at high temperature is scattered in the amorphous structure as it has short wavelength and can find imperfections as large as its wavelength. 
While at lower temperature, the mean free path of thermal phonons are determined by the sample size as the wavelength exceeds the sample size and another effect dominates the $T$-dependence of $\kappa$. 

Focusing on the deviation from the most crystallized ones (58.3\% for $C$ and 51\% for $\kappa$), amorphous contribution of PET appears as a bump in $C/T^3$ at $T\approx3$\,K and high $\kappa$ values at $T<10$\,K. 
Because the contributions have different $T$-dependence from that of 3D phonon ($\propto T^3$ for both~\cite{pobell}), 
the contribution is naturally associated with extra modes in amorphous structure, 
 and they presumably are soft localized vibrational modes recently studied for amorphous materials with theoretical approach~\cite{Mizuno14112017}.

According to Figs.~\ref{CE_HCcube_PET} and \ref{CE_TT_PET}, the crystallinity of CEs is as low as PET with 5--10\% crystallinity. 
It is consistent with previous X-ray measurements of TA (p.52 of Ref.~\cite{CTCER}) saying ``amorphous with sign of crystallinity''. 
Beside the bump and plateau features, the absolute values of $C$ and $\kappa$ of CEs are also similar to those of PET. 
This indicates that vibrational properties of CEs and PET are very similar in wide range of energy scales. 
As a side note, the extra contribution of $C$ at $T<1$\,K due to the structural rearrangements described in Section~\ref{sec_sh} appears as an upturn in Fig.~\ref{CE_HCcube_PET}, which does not have clear dependence on the crystallinity because it is sensitive only to atomic-scale arrangements. 

\subsection{Chemical structure of BT resins}
Blends of different monomers usually unite together and form a copolymer.  
BT resins, blends of CEs and bismaleimides (BMIs), were first postulated to coreact during curing, but now they are believed to have interpenetrating polymer networks of CE and BMI, instead of cocured matrix, because there are few signs of copolymerization~\cite{CTCER}. 

\begin{figure}[t!]
\centering
\includegraphics[width=0.85\columnwidth]{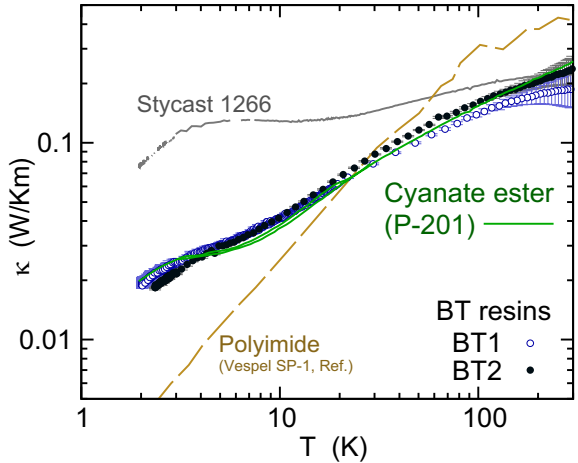}
\caption{Thermal conductivity of BT resins compared to CE (P-201), EP (Stycast 1266), and polyimide (Vespel SP-1)~\cite{doi:10.1063/1.3292433}.}
\label{BT_TT}
\end{figure}

\begin{figure}[t!]
\centering
\includegraphics[width=0.97\columnwidth]{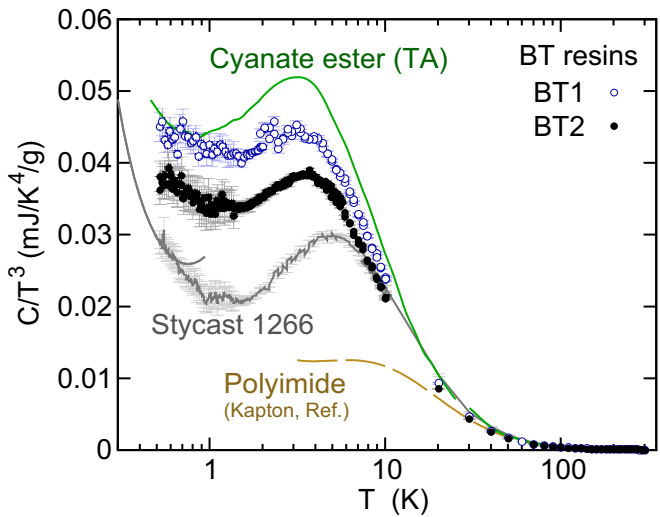}
\caption{Specific heat over temperature cube of BT resins compared to CE (TA), EP (Stycast 1266, low-temperature data is from Ref.~\citenum{pobell}), and polyimide (Kapton)~\cite{nist}.}
\label{BT_HC_cube}
\end{figure}

Figure~\ref{BT_TT} shows $\kappa$ of BT resins compared with P-201 (CE), Stycast 1266 (EP), and Vespel SP-1 (polyimide). 
BT resins have similar values of $\kappa$ to CEs, which indicates that the basic skeleton of CEs (folded triazine rings) remains in the BT resins.  
However, the characteristic plateau at $T=$3--7\,K almost disappears for BT resins, especially for BT2. 
At the same time, the characteristic bump in $C/T^3$ at $T\approx3$\,K also gets lower in BT resins, as shown in Fig.~\ref{BT_HC_cube}. 
Thus BT resins appear to have less amount of amorphous component than CEs. 
It indicates that additive BMI enters into the free space or sparse amorphous structure of CEs and forms interpenetrating networks. 
It is consistent with the fact that BT2 with simple side chains (only \ce{-H}) has stronger effects than BT1 with \ce{-Me} and \ce{-Et}. 
This is also consistent with that BT resins have less free volume ($\approx$1.7\%~\cite{PAT:PAT1898}) than CEs, often characterized by their large free volume ($\approx$2.2\%~\cite{Zeng2010966}). 

Additionally, we discuss the crystallinity of polyimide and Stycast 1266 compared in the figures.
For polyimide, amorphous contribution is hardly found in either Figs.~\ref{BT_TT} or \ref{BT_HC_cube}. 
It is natural because thermoplastic polyimide is stretched in the fabrication process and the polymer chains are aligned. 
For Stycast 1266, the $\kappa$ in Fig.~\ref{BT_TT} shows a large plateau at 4--20\,K. 
On the other hand, the bump of $C$ at $T\approx5$\,K in Fig.~\ref{BT_HC_cube} is not so high compared to that of CEs at $T\approx3$\,K. 
As the transport property ($\kappa$) is sensitive to the connectivity and the extensive property ($C$) indicates the quantity, it suggests that amorphous regions and corresponding vibrational modes in Stycast 1266 are de-localized and contribute the thermal transportation more actively than more localized ones in CEs.

\section{Conclusions}
We measured specific heat $C$, thermal conductivity $\kappa$, and magnetic susceptibility $\chi$ of polymers of different kinds of cyanate ester resins (CEs) for future  use of them as a glue and casting resin compatible at wide temperature range and/or in high magnetic field. 
The results show that CEs have small $C$ and $\kappa$ at low temperature, and negligible $\chi$, as expected for pure polymers. 
Along with their other advantages of high glass transition temperature ($\approx300$\,\degree C), low outgassing properties, and absence of health problems, CEs can be an excellent alternative to Stycast 1266 in some situations. 
Compared to other materials, CEs are characterized by low thermal diffusivity ($\equiv \kappa/\rho C$) throughout the temperature range of 2--300\,K. 
It indicates that CEs can be used to filter out rapid temperature fluctuations. 
Although the thermal properties below 30\,K are not dependent on the side chains, there would be some scope to tailor the side chains to make it more convenient to work with. 
Altering the side chains are known to affect the melt viscosity, glass transition temperature, mechanical toughness, and color~\cite{CTCER}. 
Blending with bismaleimide reduces the amorphous contribution in the thermal properties which appears at $T<10$\,K. 

Along with the practical purposes for thermal design, thermal measurements at low temperature can provide structural information through the eyes of thermal phonon whose wavelength varies from nanometers to submillimeters, and of atoms tunneling between possible positions at low temperature. 
From such analysis, the structure of CEs is almost amorphous over a wide range of length scales, but in atomic scales, each atom has few possible positions to tunnel around. 
Similar analysis on BT resins, blends of CEs and BMIs, indicates that CEs keep the main structure (triazine rings) of the characteristic polymer network even in BT resins and that BMIs penetrate through the amorphous regions in the CE network and increase the crystallinity of the final products. 

\section{Acknowledgements}
The authors would like to thank Dr. Ryo Toda for technical assistance with the experiments. We also thank Dr. David T. Peat for providing language help. 

\bibliography{CEbib_withLTpro}

\end{document}